\documentclass[runningheads]{llncs}
\usepackage[T1]{fontenc}
\usepackage{graphicx}

\usepackage{booktabs}
\usepackage{amsmath}
\usepackage{amssymb}
\usepackage{multirow}
\usepackage[table]{xcolor}
\usepackage{mathtools}
\usepackage{algorithm}
\usepackage[noend]{algpseudocode}
\usepackage{setspace}
\usepackage{lipsum}

\begin{document}

\title{Simulation-Based Segmentation of\\ Blood Vessels in Cerebral 3D OCTA Images}
\titlerunning{Simulation-Based Segmentation of Cerebral 3D OCTA Images}

\author{
Bastian Wittmann\inst{1}\and
Lukas Glandorf\inst{2,5}\and
Johannes C. Paetzold\inst{7}\and \\
Tamaz Amiranashvili\inst{1,6}\and
Thomas Wälchli\inst{3,4,8,9}\and \\
Daniel Razansky\inst{2,5}\and
Bjoern Menze\inst{1}
}

% \author{
% Thomas Wälchli\inst{1,2,3,4}
% }

% \institute{
% Neuroscience Center Zurich, University of Zurich, Zurich, Switzerland \and
% Krembil Research Institute, Toronto Western Hospital, Toronto, Canada \and
% Division of Neurosurgery, University Hospital Zurich, Zurich, Switzerland \and
% Division of Neurosurgery, Toronto Western Hospital, Toronto, Canada
% }

\authorrunning{B. Wittmann et al.}

\institute{
Department of Quantitative Biomedicine, University of Zurich, Switzerland \\
\email{\{bastian.wittmann,bjoern.menze\}@uzh.ch} \and
Institute of Pharmacology and Toxicology, University of Zurich, Switzerland\and
Neuroscience Center Zurich, University of Zurich, Switzerland\and
Division of Neurosurgery, University Hospital Zurich, Switzerland\and
Institute for Biomedical Engineering, ETH Zurich, Switzerland\and
Department of Computer Science, Technical University of Munich, Germany\and
Department of Computing, Imperial College London, England\and
Krembil Research Institute, Toronto Western Hospital, Canada\and
Division of Neurosurgery, Toronto Western Hospital, Canada
}

\maketitle              % typeset the header of the contribution

% \begin{figure}[h]
% \vspace{-0.5em}
% \centerline{\includegraphics[width=0.8\linewidth]{imgs/synthetic_renderings.jpg}}
% \caption{3D renderings of our proposed synthetic cerebral 3D OCTA images.}
% \label{fig:synthetic_renderings}
% \vspace{-2.5em}
% \end{figure}

\begin{abstract}
\setcounter{footnote}{0}
Segmentation of blood vessels in murine cerebral 3D OCTA images is foundational for \textit{in vivo} quantitative analysis of the effects of neurovascular disorders, such as stroke or Alzheimer's, on the vascular network. However, to accurately segment blood vessels with state-of-the-art deep learning methods, a vast amount of voxel-level annotations is required. Since cerebral 3D OCTA images are typically plagued by artifacts and generally have a low signal-to-noise ratio, acquiring manual annotations poses an especially cumbersome and time-consuming task. To alleviate the need for manual annotations, we propose utilizing synthetic data to supervise segmentation algorithms. To this end, we extract patches from vessel graphs and transform them into synthetic cerebral 3D OCTA images paired with their matching ground truth labels by simulating the most dominant 3D OCTA artifacts. In extensive experiments, we demonstrate that our approach achieves competitive results, enabling annotation-free blood vessel segmentation in cerebral 3D OCTA images.

\end{abstract}

\section{Introduction and Motivation}
% what is octa, how does it work, what does it provide, what is it good for in medicine
Optical Coherence Tomography Angiography (OCTA) is a high-resolution imaging technique that captures blood vessels by detecting flow-induced temporal changes of the backscattered signal caused by red blood cell (RBC) movement. This enables OCTA to provide \textit{in vivo}, three-dimensional (3D) images of blood vessels, including minuscule capillaries, in tandem with blood flow information (when paired with Doppler OCT). The combination of high-resolution, \textit{in vivo} vascular images with their corresponding blood flow mapping enables researchers to monitor brain vascular dynamics~\cite{walchli2023shaping} and thus gain valuable, unique insights into neurovascular disorders such as stroke~\cite{erdener2019spatio} or Alzheimer’s~\cite{walek2023near}.

% why segmentation
The automated analysis of 3D OCTA images typically builds on an initial blood vessel segmentation stage~\cite{stefan2020deep,walek2023near}, which lays the foundation for advanced vasculature analysis.
Currently, blood vessel segmentation is almost exclusively performed by supervised deep-learning methods, relying on large, manually annotated datasets curated by trained experts. In the context of 3D OCTA images, however, acquiring manual annotations poses an especially cumbersome task due to dominant imaging artifacts~\cite{li2022blood,zhu2020visibility,hormel2021artifacts} and the additional complexity introduced by the requirement of 3D voxel-level consistent annotations of densely connected capillaries. 
Furthermore, high variability in OCT system design and acquisition protocols limits the use of annotated data from different OCT setups, as supervised methods fail to generalize across these variations. Therefore, most methods refrain from analyzing 3D OCTA images and instead focus on 2D \textit{en-face} projections, discarding 3D information relevant for a more comprehensive analysis.

% what are we doing
In light of the absence of large-scale annotated cerebral 3D OCTA datasets and to address the challenge of high variability in OCTA image characteristics, we propose a synthesis pipeline that can be adapted to the data at hand with little effort to generate a vast amount of synthetic data. To be precise, we first transform vessel graphs derived from real 
murine vasculature~\cite{walchli2021hierarchical} into voxelized volumes, maintaining relevant morphological properties. As a second step, we modify these voxelized volumes by simulating the most dominant 3D OCTA image acquisition artifacts. This results in realistic, synthetic cerebral 3D OCTA images paired with their intrinsically matching ground truth labels given by the unmodified voxelized volumes.
Subsequently, we leverage our generated synthetic dataset to train a segmentation network, which allows us to essentially erase the need for manual annotations while ensuring accurate segmentation maps.

\noindent
Our core contributions are summarized in brief as follows:
\begin{enumerate}
\itemsep0em 
    \item We address the lack of manual annotations by generating a vast amount of synthetic cerebral 3D OCTA images with matching ground truth labels.

    \item We identify projection artifacts, angle-dependent signal loss, and local granular noise patterns as the most dominant artifacts in cerebral 3D OCTA images and model them adequately in our simulation.
    
    \item We demonstrate that our simulation-based segmentation approach enables accurate, annotation-free segmentation of cerebral 3D OCTA images.

    \item We tackle high variations in OCT system design and acquisition protocols by proposing a synthesis pipeline that can be adapted to the data at hand.

    \item We open-source our code, synthetic dataset, and manually annotated OCTA images to serve as a benchmark for cerebral 3D OCTA segmentation.\footnote{https://github.com/bwittmann/syn-cerebral-octa-seg}
\end{enumerate}

\begin{figure}[t]
\centerline{\includegraphics[width=0.8\linewidth]{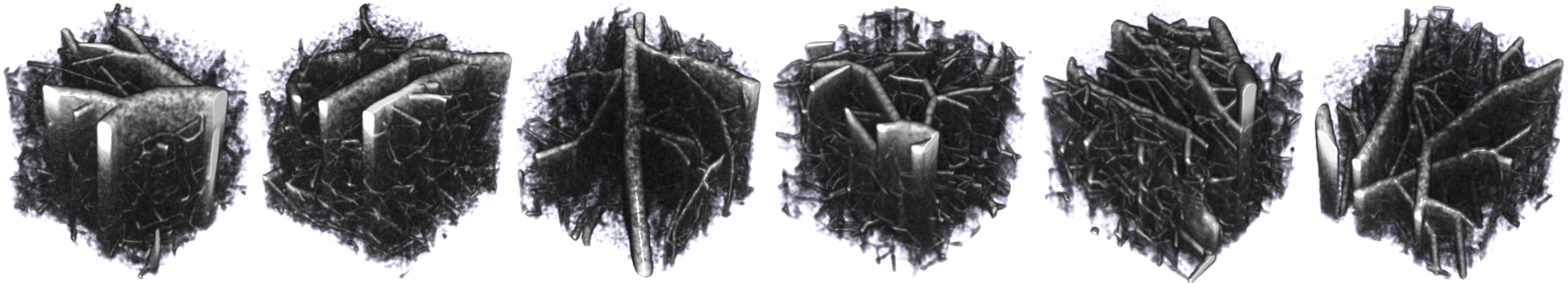}}
\caption{3D renderings of our proposed synthetic cerebral 3D OCTA images.}
\label{fig:synthetic_renderings}
\end{figure}

\section{Related Works}
\subsubsection{3D OCTA Segmentation:}
Due to the lack of manual annotations, 3D OCTA segmentation remains largely unexplored~\cite{meiburger2021automatic}. In the context of the more extensively studied retinal OCTA images, some work experimented with 3D-to-2D segmentation~\cite{li2020image,li2020ipn,wu2021paenet}, 2D-to-3D segmentation~\cite{yu20213d,yu2021cross}, and unsupervised 3D-to-3D segmentation~\cite{hu2021life}, relying on an auxiliary capillary-enhanced modality. Since cerebral 3D OCTA images differ significantly from retinal 3D OCTA images (see Suppl., Fig.~\ref{fig:retina_vs_cerebral}), leveraging techniques tailored to retinal OCTA images is practically impossible. To address this issue, Stefan et al.~\cite{stefan2020deep,walek2023near} recently opted to train a 3D CNN on a manually annotated cerebral 3D OCTA volume to segment murine cerebral vasculature. However, their CNN is trained on a volume acquired by their in-house OCT system and, therefore, does not generalize well to OCTA images from different setups (see Table~\ref{tab:quantitative_results}, a). In this work, we eliminate this issue by proposing to train on synthetic cerebral 3D OCTA images that can be tuned to match characteristics of OCTA images at hand (see Suppl., Fig.~\ref{fig:sim_params}).

\subsubsection{Simulation-Based Blood Vessel Segmentation:}
Over the last few years, synthetic images have been successfully utilized to train blood vessel segmentation algorithms in various 2D imaging modalities~\cite{ma2021iccv,shi2023freecos}. In the realm of OCTA, Menten et al.~\cite{menten2022physiology} introduced a physiology-based simulator with the aim of creating synthetic retinal 2D \textit{en-face} projections. Building upon this idea, Kreitner et al.~\cite{kreitner2024synthetic} simulated retinal vasculature more accurately, resulting in improved synthetic retinal 2D \textit{en-face} projections. However, none of the above methods are tailored to 3D OCTA images and their unique artifacts. Furthermore, these methods rely on artificial fractals~\cite{ma2021iccv,shi2023freecos} or retinal vessel simulators~\cite{menten2022physiology,kreitner2024synthetic} prone to errors to generate ground truth labels. In contrast, our work is specifically developed for cerebral 3D OCTA images and relies on real angioarchitectural properties preserved in vessel graphs derived from murine vasculature.

\section{Method}
\begin{figure}[h]
\centerline{\includegraphics[width=\linewidth]{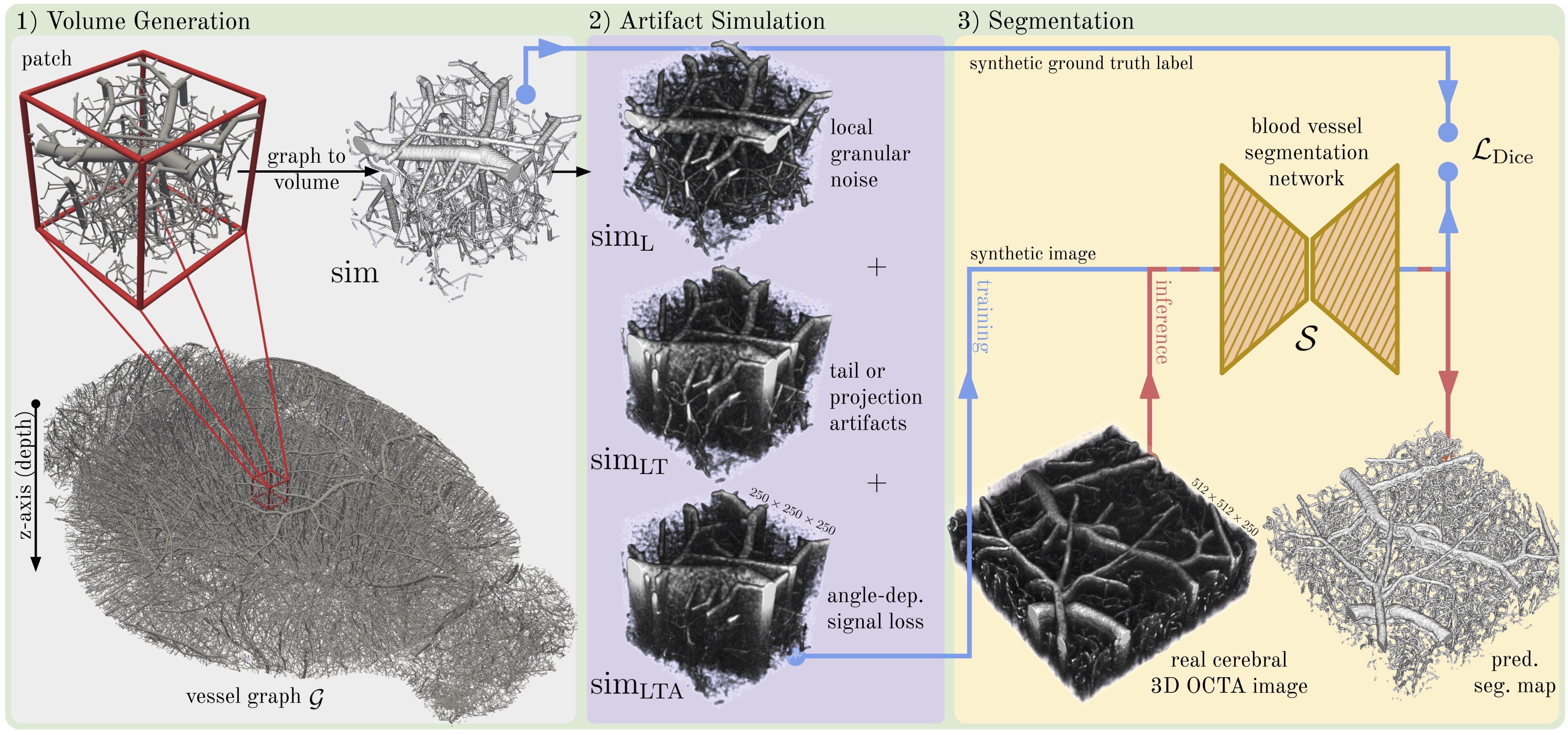}}
\caption{Overview of our proposed method. First, we extract patches from vessel graphs and transform them into a vast amount of voxelized volumes; second, we transform the voxelized volumes into synthetic cerebral 3D OCTA images by simulating the most dominant image acquisition artifacts; and third, we use our synthetic cerebral 3D OCTA images paired with their matching ground truth labels to train a segmentation network.}
\label{fig:method}
\end{figure}

Our proposed method is divided into three stages: volume generation, artifact simulation, and segmentation (see Fig.~\ref{fig:method}). The three stages are described below.

\subsubsection{1) Volume Generation:}
% introduction to vessel graphs
The volume generation stage relies on vessel graphs $\mathcal{G}\coloneqq(\mathcal{V}, \mathcal{E})$, comprised of nodes $\mathcal{V}$ representative of vessel branching points or locations with stronger curvature and edges $\mathcal{E}$ representative of blood vessels. While nodes are defined by $x$-, $y$-, and $z$-coordinates, edges
are spanned between two nodes and contain solely information regarding vessel radii $r$. Therefore, blood vessels appear tubular. In this work, we make use of graph representations of murine whole-brain vascular corrosion casts~\cite{walchli2021hierarchical}, which accurately preserve cerebral vasculature (see Fig.~\ref{fig:method}) all the way down to the smallest capillaries.

\begin{figure}[t]
\centerline{\includegraphics[width=0.9\linewidth]{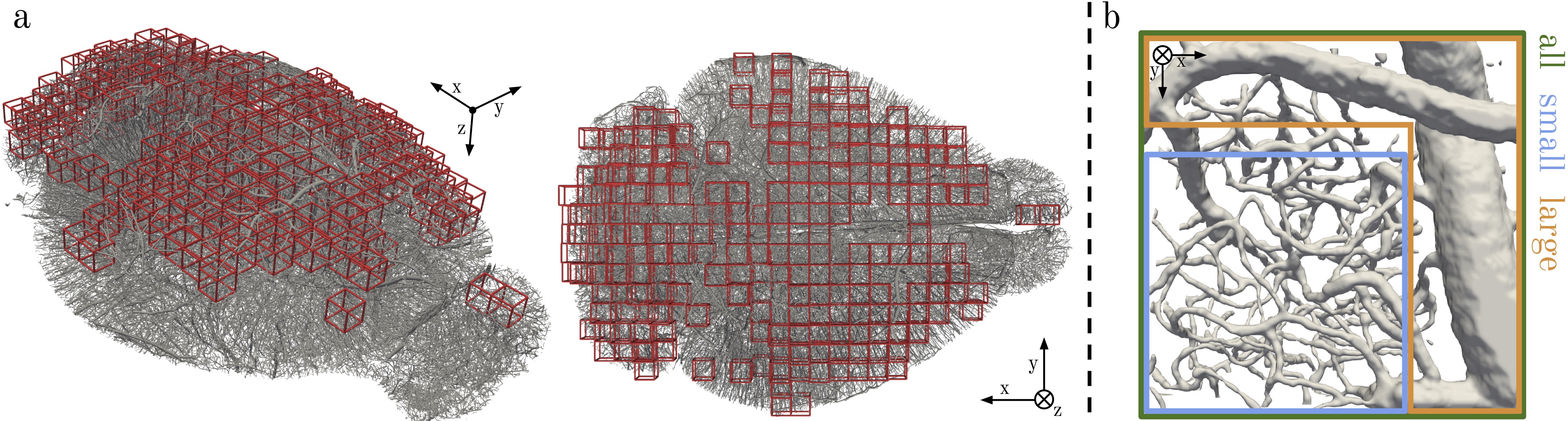}}
\caption{a) Visualization of sampled patches in a graph representation of a vascular corrosion cast; b) exemplary manual annotation, including regions provided to determine vessel size-specific segmentation performance (all, small, large).}
\label{fig:patches_size}
\vspace{-0.5em}
\end{figure}

% patch extraction and graph2volume.
We sample the vascular corrosion casts in a grid-like manner (see Fig.~\ref{fig:patches_size}, a) to extract patches, building the foundation for our simulation. We pay special attention to exclusively sample patches that match the FOV and the characteristics of vasculature contained in cerebral 3D OCTA images. Specifically, we discard patches based on the following criteria: 1) the patch originates from brain regions difficult to image with modern OCT systems (depth > 3 mm); 2) the patch is sparsely populated (less than 2,000 vessels); and 3) the patch contains no larger vessels ($r$ > 13 \textmu m). To transform the extracted vessel graph patches to voxelized volumes, we plot the centerlines of vessels and subsequently perform 3D binary morphological dilation. We additionally store metadata for each voxel, being the radius $r$ and the smallest angle to the z-axis $\theta_{\text{z}}$ of the originating vessel. In total, we generate 1,137 voxelized volumes of isotropic voxel size (2 \textmu m) and shape $250 \times 250 \times 250$ from six whole-brain vascular corrosion casts.

\begin{figure}[h]
\centerline{\includegraphics[width=0.9\linewidth]{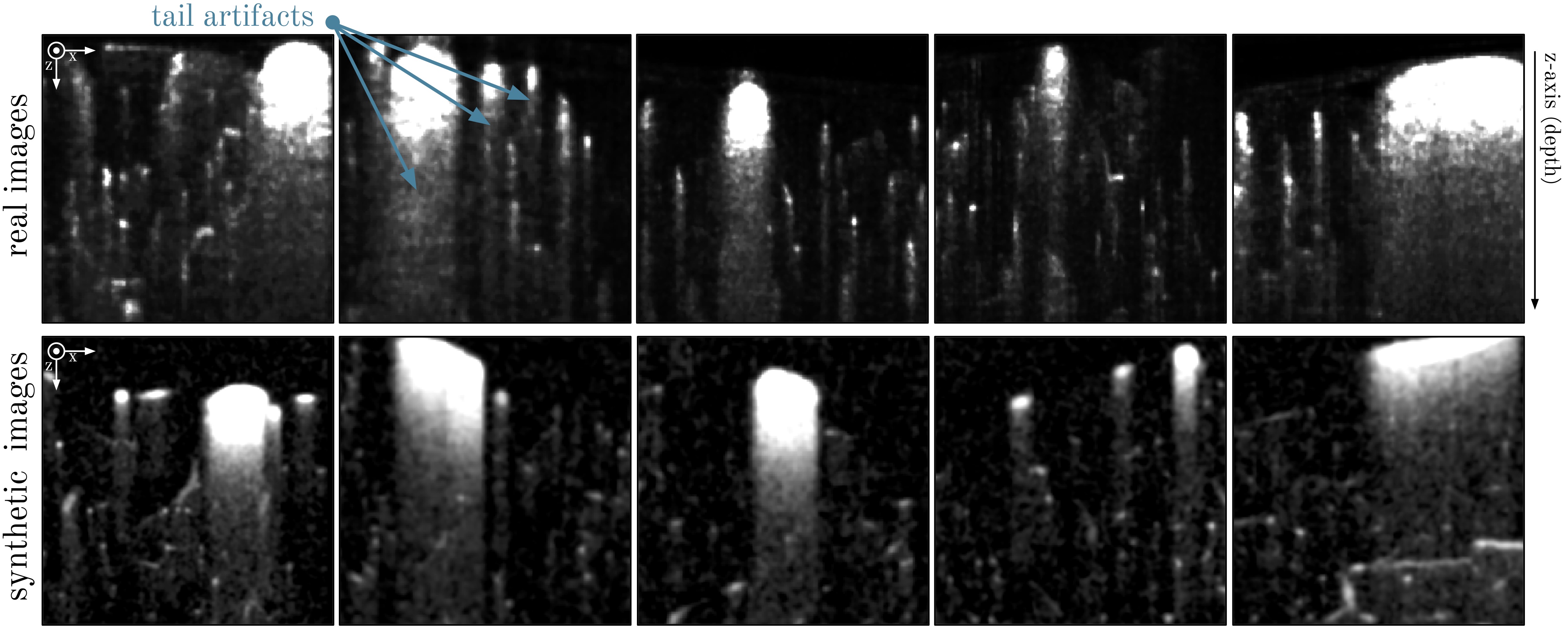}}
\caption{Slices of real (top) and synthetic (bottom) cerebral 3D OCTA images. It should be highlighted that we accurately match 3D OCTA-specific artifacts, resulting in synthetic images almost indistinguishable from real images.}
\label{fig:synthetic_bscan}
\end{figure}

\subsubsection{2) Artifact Simulation:}
We identify projection artifacts, angle-dependent signal loss, and random local granular noise patterns as the three most dominant artifacts in cerebral 3D OCTA images and simulate them in our generated voxelized volumes. In the following, we briefly address these artifacts and describe how we model them in our algorithmic implementation (see Suppl., Algo.~\ref{alg:artifact_simulation}). It should be highlighted that our simulation relies solely on a few parameters, which can be adjusted in minutes to match the characteristics of OCTA images at hand (see Suppl., Fig.~\ref{fig:sim_params}). Slices of our generated synthetic cerebral 3D OCTA images are depicted in Fig.~\ref{fig:synthetic_bscan}, while 3D renderings can be found in Fig.~\ref{fig:synthetic_renderings}.

\textit{Projection (or Tail) Artifacts:}
To accurately capture the depth of blood vessels, OCTA relies on backscattered photons from single scattering events with RBCs. However, multiple scattering (forward and backward) of an incident photon with multiple RBCs and deeper tissue layers artificially elongates the photon path length and thus results in incorrect depth estimates appearing as artifacts beneath blood vessels (see Fig.~\ref{fig:synthetic_bscan}). These artifacts are referred to as projection or tail artifacts~\cite{li2022blood,hormel2021artifacts} and limit accurate vasculature quantification, as they obscure the signal beneath large pial vessels and distort the tubular appearance of blood vessels. We simulate projection artifacts (see Algo.~\ref{alg:artifact_simulation}, line~16) as an exponential signal decay modeled by a geometric progression, reflecting the nature of the multiple scattering process~\cite{hormel2021artifacts}. The amount of involved scattering interactions and experienced multiple scattering events depends on the local concentration and distribution of RBCs. Therefore, we derive the length of projection artifacts and their initial intensities primarily from the radius $r$ of the respective vessels. We finally add random Gaussian noise onto the simulated tail artifacts to match the stochastic behavior of multiple scattering more accurately.

\textit{Angle-Dependent Signal Loss:}
In 3D OCTA images, signal from microvessels, such as capillaries, depends strongly on their angular orientation. Studies have shown that this can be attributed to the longitudinal elongation of RBCs~\cite{secomb2001motion} in microvessels in conjunction with the orientation dependence of RBC backscattering. To be precise, longitudinally elongated RBCs flowing in parallel to the incident photon (or the z-axis) present a drastically smaller effective scattering cross-section to the photon compared to orthogonally flowing longitudinally elongated RBCs. The reduction in effective scattering cross-section thus renders microvessels running in parallel to the light beam almost invisible in OCTA images~\cite{zhu2020visibility}. To account for this angle-dependent signal loss, we exponentially decay signal depending on how much the angle between the blood vessel and the z-axis $\theta_{\text{z}}$ deviates from $90^{\circ}$ (see Algo.~\ref{alg:artifact_simulation}, line~8). We base the exponential signal decay on an experimental study from Zhu et al.~\cite{zhu2020visibility}, investigating the relationship between RBC backscattering ratio and vessel orientation. In our simulation, a sigmoid function acts as a soft threshold between micro- and macrovessels. The smooth transition prevents the abrupt emergence of angle-dependent signal loss.

\textit{Local Granular Noise Patterns:}
Further, we aim to match local granular noise patterns and intensity variations, primarily arising from weak residual and sub-cellular motion~\cite{munter2020dynamic}, spontaneous neuronal activation~\cite{tang2021imaging}, and system noise. To this end, we add Gaussian noise to the whole synthetic image, followed by Gaussian smoothing (see Algo.~\ref{alg:artifact_simulation}, line~21).

\subsubsection{3) Segmentation:}
The 1,137 generated synthetic cerebral 3D OCTA images are utilized together with the unmodified, binary voxelized volumes, acting as ground truth labels (see Fig.~\ref{fig:method}, blue), to train an off-the-shelf deep learning-based segmentation network $\mathcal{S}$ in a supervised manner. During inference, the trained segmentation network is applied to real murine cerebral 3D OCTA images.

\section{Experiments and Results}
To evaluate our method, a trained expert manually annotated six cerebral 3D OCTA volumes of isotropic voxel size (2 \textmu m) and shape $160 \times 160 \times 160$ over the course of six months. The OCTA volumes originate from multiple experiments involving various mice and are acquired by our OCT setup (similar to~\cite{marchand2018imaging}). We split annotations into three test volumes, one validation volume, and two volumes used to train the upper bound (see Table~\ref{tab:quantitative_results}, a). Since 3D OCTA images contain vessels of various sizes that are affected to different degrees by size-specific artifacts, we provide regions containing primarily micro- (small) and macrovessels (large) (see Fig.~\ref{fig:patches_size}, b) to determine size-specific segmentation performance (see Table~\ref{tab:quantitative_results}). We report Dice and topology-aware centerline Dice (clDice)~\cite{shit2021cldice} scores to measure the algorithms' ability to accurately preserve tubular appearance and vessel connectivity. We opt for the 3D U-Net architecture~\cite{isensee2021nnu} to present the segmentation network $\mathcal{S}$ and tune its parameters on the validation volume\footnote{For detailed information, please refer to the code (\texttt{config.yaml}).}.

First, we compare the performance of the U-Net trained on our proposed synthetic data (see Table~\ref{tab:quantitative_results}, c) to the same U-Net trained on real, annotated data (see Table~\ref{tab:quantitative_results}, a) and to traditional, annotation-free segmentation techniques~\cite{frangi1998multiscale,otsu1979threshold} (see Table~\ref{tab:quantitative_results}, b). Analysis of quantitative results leads to the conclusion that the U-Net trained on our synthetic data accurately segments vasculature in 3D OCTA images, achieving competitive results without the need for manual annotations. The U-Net trained on our synthetic data not only outperforms traditional techniques in all metrics but also a U-Net trained on a real, annotated volume of shape $154 \times 154 \times 77$ obtained from a different OCT setup~\cite{stefan2020deep}. Even though we ensured that important properties, such as the voxel size and the volume's intensity range, match the images acquired by our OCT setup, the U-Net trained on data from \cite{stefan2020deep} demonstrates poor generalization. By adjusting the simulation parameters used to generate our synthetic images to data acquired by our OCT setup, we almost match the performance of the upper bound. The observed difference to the upper bound is in line with related works~\cite{ma2021iccv,shi2023freecos,kreitner2024synthetic}.

\begin{table}[t]
\centering
\scriptsize
\caption{Quantitative results achieved on the test set: a) U-Net trained on real, manually annotated data; b) traditional techniques; c) U-Net trained on our synthetic 3D OCTA data; d) ablation on simulated artifacts; e) ablation on curvature; f) ablation on vessel graphs. We report mean and std values based on four random seeds (0 - 3).}
\label{tab:quantitative_results}
\vspace{-0.5em}
\begin{tabular}{p{10pt} p{40pt} l|c c|c c|c c}
\toprule
\multirow{2}{*}{} & \multirow{2}{*}{Method} & \multirow{2}{*}{Data} 
& \multicolumn{2}{c|}{all}& \multicolumn{2}{c|}{small}& \multicolumn{2}{c}{large}\\
&&& $\text{Dice}\uparrow$ & $\text{clDice}\uparrow$ & $\text{Dice}\uparrow$ & $\text{clDice}\uparrow$ & $\text{Dice}\uparrow$ & $\text{clDice}\uparrow$\\ 

\midrule    % trained on real data
\multirow{2}{*}{a}
& 3D U-Net & real (ours)* & \cellcolor{teal!40}79.46\tiny{±0.18} & \cellcolor{teal!40}85.29\tiny{±0.16} & \cellcolor{teal!40}74.31\tiny{±0.08} & \cellcolor{teal!40}87.86\tiny{±0.12} & \cellcolor{teal!40}82.54\tiny{±0.25} & \cellcolor{teal!40}70.93\tiny{±0.21}\\
& 3D U-Net & real (\cite{stefan2020deep}) & 54.13\tiny{±1.37} & 65.41\tiny{±1.09} &52.78\tiny{±0.90} & 67.57\tiny{±1.16} & 55.07\tiny{±2.17} & 48.50\tiny{±1.39}\\

\midrule    % traditional methods
\multirow{2}{*}{b}
& Frangi & - & 40.84\tiny{±0.00} & 53.13\tiny{±0.00} & 58.52\tiny{±0.00} & 65.96\tiny{±0.00} & 19.57\tiny{±0.00} & 30.95\tiny{±0.00}\\
& Otsu & - & 50.62\tiny{±0.00} & 33.47\tiny{±0.00} & 42.99\tiny{±0.00} & 49.34\tiny{±0.00} & 51.63\tiny{±0.00} & 11.20\tiny{±0.00}\\

\midrule    % synthetic data
\multirow{1}{*}{c}
& 3D U-Net & syn. ($\text{sim}_{\text{LTA}}$) & \cellcolor{teal!20}74.83\tiny{±0.23} & \cellcolor{teal!20}80.92\tiny{±0.13} & \cellcolor{teal!20}66.90\tiny{±0.16} & \cellcolor{teal!20}81.27\tiny{±0.14} & \cellcolor{teal!20}80.66\tiny{±0.36} & \cellcolor{teal!10}69.19\tiny{±0.31}\\ 

\midrule    % synthetic data ablations
\multirow{3}{*}{d}
& 3D U-Net & syn. ($\text{sim}_{\text{}}$) & 50.85\tiny{±0.88} & 29.98\tiny{±1.15} & 54.26\tiny{±1.77} & 62.63\tiny{±2.01} & 47.33\tiny{±1.27} & 9.87\tiny{±0.99}\\
& 3D U-Net & syn. ($\text{sim}_{\text{L}}$) & 52.68\tiny{±1.18} & 46.15\tiny{±2.15} & 56.67\tiny{±0.99} & 61.34\tiny{±1.13} & 47.61\tiny{±1.59} & 28.77\tiny{±2.39}\\ 
& 3D U-Net & syn. ($\text{sim}_{\text{LT}}$) & 70.38\tiny{±0.43} & 72.80\tiny{±0.73} & 57.50\tiny{±0.93} & 73.60\tiny{±0.53} & 79.81\tiny{±0.30} & 59.48\tiny{±0.75}\\

\midrule    % ablations on curvature
\multirow{1}{*}{e}
& 3D U-Net & syn. ($\text{sim}_{\text{LTAC}}$) & \cellcolor{teal!10}74.46\tiny{±0.19} & \cellcolor{teal!10}80.84\tiny{±0.19} & 66.50\tiny{±0.21} & 80.75\tiny{±0.11} & \cellcolor{teal!10}80.33\tiny{±0.41} & \cellcolor{teal!20}69.55\tiny{±0.75}\\

\midrule    % ablations on vessel graph
\multirow{1}{*}{f}
& 3D U-Net & syn. ($\text{sim}_{\text{LTA}}^{\text{SAT}}$) & 60.00\tiny{±1.28} & 74.48\tiny{±1.10} & \cellcolor{teal!10}66.77\tiny{±0.53} & \cellcolor{teal!10}80.88\tiny{±0.19} & 51.84\tiny{±3.86} & 52.92\tiny{±1.82}\\

\bottomrule
\multicolumn{9}{l}{\tiny{$^{*}$Upper bound trained on our annotated images. Annotating 2 training volumes consumed $\sim$2 months.}}
\end{tabular}
\end{table}

% ablation of artifacts
Further, we investigate the influence of simulated artifacts by sequentially including local granular noise patterns ($\text{L}$), tail artifacts ($\text{T}$), and angle-dependent signal loss ($\text{A}$) in our synthetic images (see Table~\ref{tab:quantitative_results}, d \& c). Corresponding 3D renderings can be found in Fig.~\ref{fig:method}. We find that our proposed simulated artifacts result in substantial individual performance increases. Introducing tail artifacts, which are especially prominent in large pial vessels, \emph{e.g.}, results in a more pronounced increase with regard to large vessels, while angle-dependent signal loss mostly affects microvessels and hence boosts segmentation performance on small vessels.
% ablation of curvature
An additional ablation study investigates the importance of curvature (C) modeled by elastic deformation (see Table~\ref{tab:quantitative_results}, e). Even though we hypothesize that the difference in curvature (see Suppl., Fig.~\ref{fig:gt_comparison}) represents the biggest remaining mismatch between synthetic and real vessels, additional deformation does not increase segmentation performance, confirming the validity of modeling vessels as tubular structures.
% ablation of vessel graphs
Finally, we evaluate the influence of the underlying vessel graphs. To this end, we exchange the vascular corrosion casts originating from real murine vasculature with simulated arterial trees (SAT)~\cite{schneider2012tissue} (see Table~\ref{tab:quantitative_results}, f \& Suppl., Fig.~\ref{fig:gt_comparison}). Given that simulated arterial trees fail to match morphological properties of cerebral vasculature by, \emph{e.g.}, not containing larger pial vessels, segmentation performance diminishes, particularly for large vessels.\\

\begin{figure}[t]
\centerline{\includegraphics[width=0.9\linewidth]{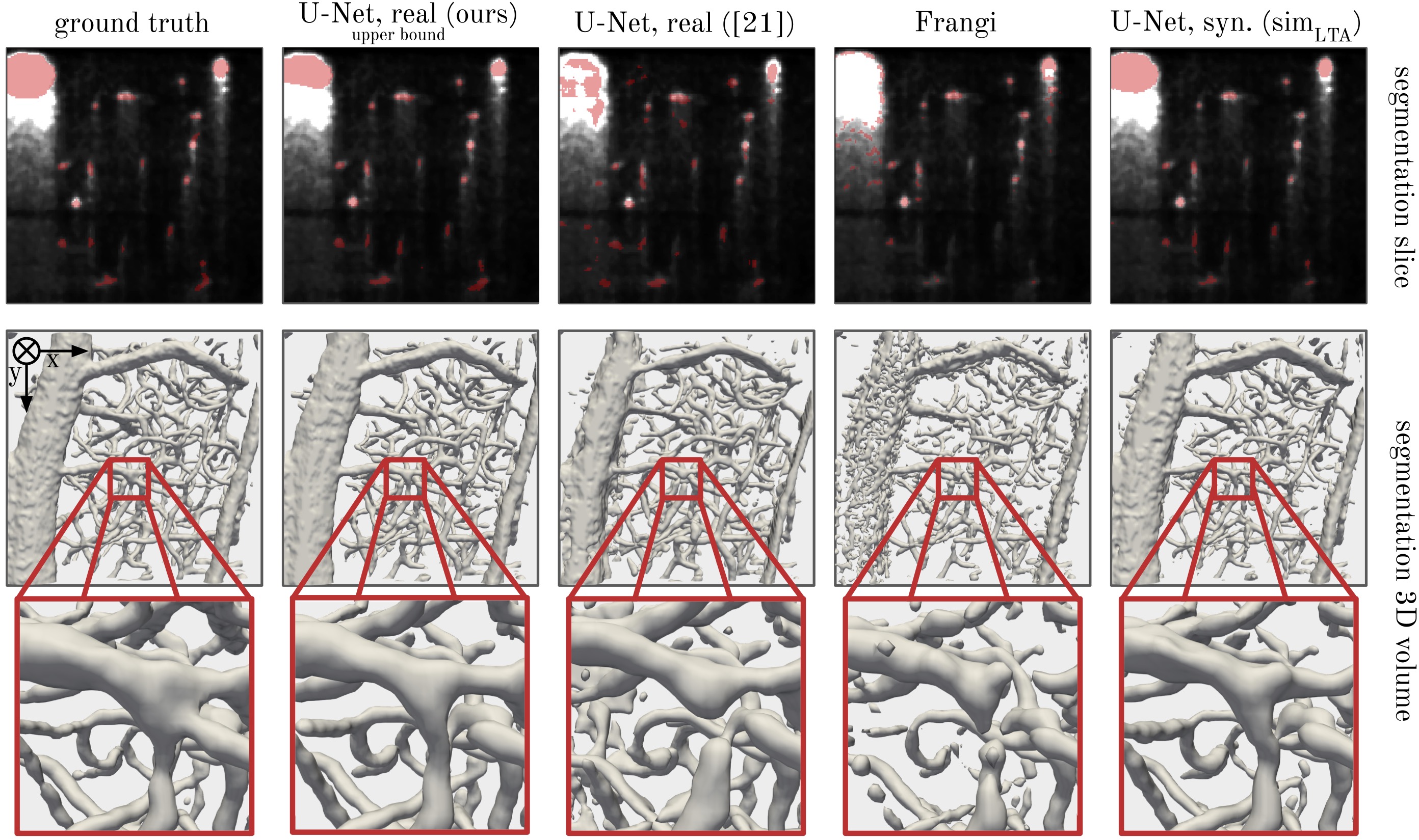}}
\caption{Qualitative results. The U-Net trained on our synthetic data (right) accurately segments vasculature, alleviating the need for labor-intensive manual annotations.}
\label{fig:qual_res}
\end{figure}

% qualitative results
Qualitative results paint a similar picture (see Fig.~\ref{fig:qual_res}). However, we hypothesize that the U-Net trained on synthetic data (right) may capture the shape of vessels more accurately, as our ground truth labels are derived from real vasculature and are, therefore, not susceptible to annotator-specific biases. Specifically, we find manual annotations of capillaries to be inflated (see Suppl., Fig.~\ref{fig:gt_comparison}), which may explain the discrepancy between quantitative and qualitative results, especially for small vessels (see Table~\ref{tab:quantitative_results}, a \& c). Analysis of Fig.~\ref{fig:qual_res} may support this hypothesis, as capillaries segmented by our method (right) appear more slender.

\section{Conclusion}
In this work, we successfully propose the use of synthetic cerebral 3D OCTA images for blood vessel segmentation to overcome the lack of available manual annotations. We simultaneously address the challenge of high variability in OCT system design and acquisition protocols, which limits the use of annotated data from different OCT setups, by proposing a synthesis pipeline that can be adapted to the data at hand with little effort. Our proposed solution not only saves time required for the laborious manual annotation process (in our case, six months) but also erases annotator-specific biases, as the underlying ground truth of our synthetic images relies on real vasculature preserved in corrosion casts. We encourage future work to propose tailored segmentation networks~\cite{shi2023freecos} or employ graph-level postprocessing steps~\cite{wittmann2024link} to improve vessel connectivity. By open-sourcing the code, the synthetic dataset, and the manually annotated OCTA images, we hope to further push the state-of-the-art, ultimately enabling large-scale quantitative analysis of disease repercussions on the vascular network.

\bibliographystyle{splncs04}
\bibliography{mybibliography}

%%% SUPPLEMENTARY MATERIAL %%%
\appendix
\title{Supplementary Material}
\author{}
\institute{}
\maketitle
\setcounter{figure}{5}

\begin{figure}[h]
\centerline{\includegraphics[width=0.9\linewidth]{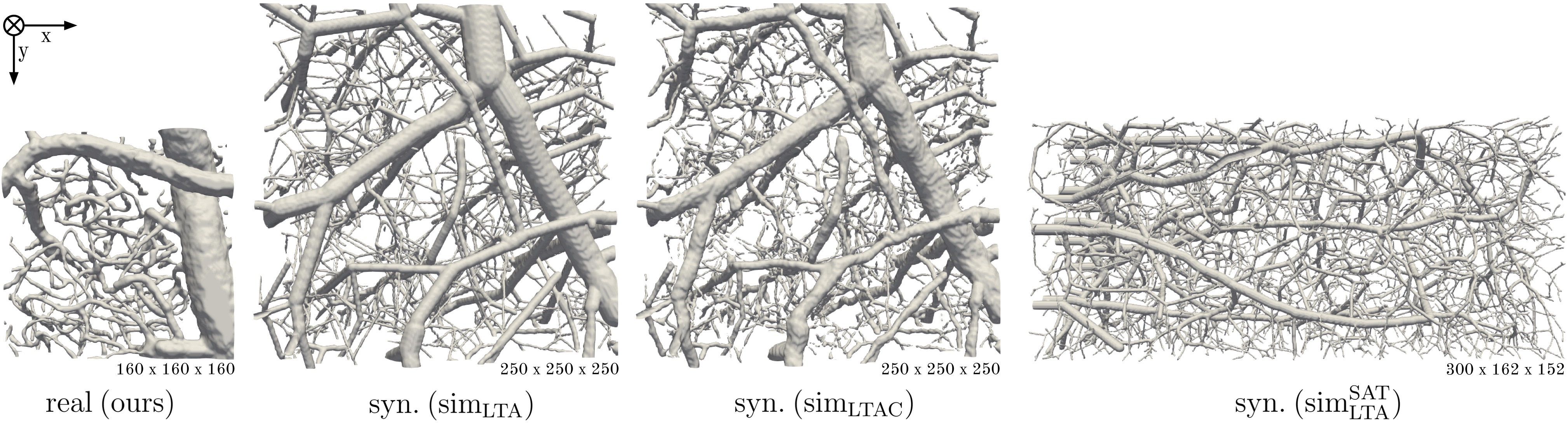}}
\caption{Comparison of ground truth labels. All labels have a similar voxel size of approximately 2 \textmu m. We compare manual annotations (left) to the ground truth labels of synthetic datasets used in our experiments. Direct comparison leads to the conclusion that capillaries appear inflated in our manual annotations, indicating an annotator-specific bias. We would like to highlight the effect of elastic deformation on curvature ($\text{sim}_{\text{LTAC}}$) and the morphological difference between labels arising from synthetic arterial trees ($\text{sim}_{\text{LTA}}^{\text{SAT}}$) and vascular corrosion casts ($\text{sim}_{\text{LTA}}$).}
\label{fig:gt_comparison}
\end{figure}

\begin{figure}[h]
\centerline{\includegraphics[width=0.9\linewidth]{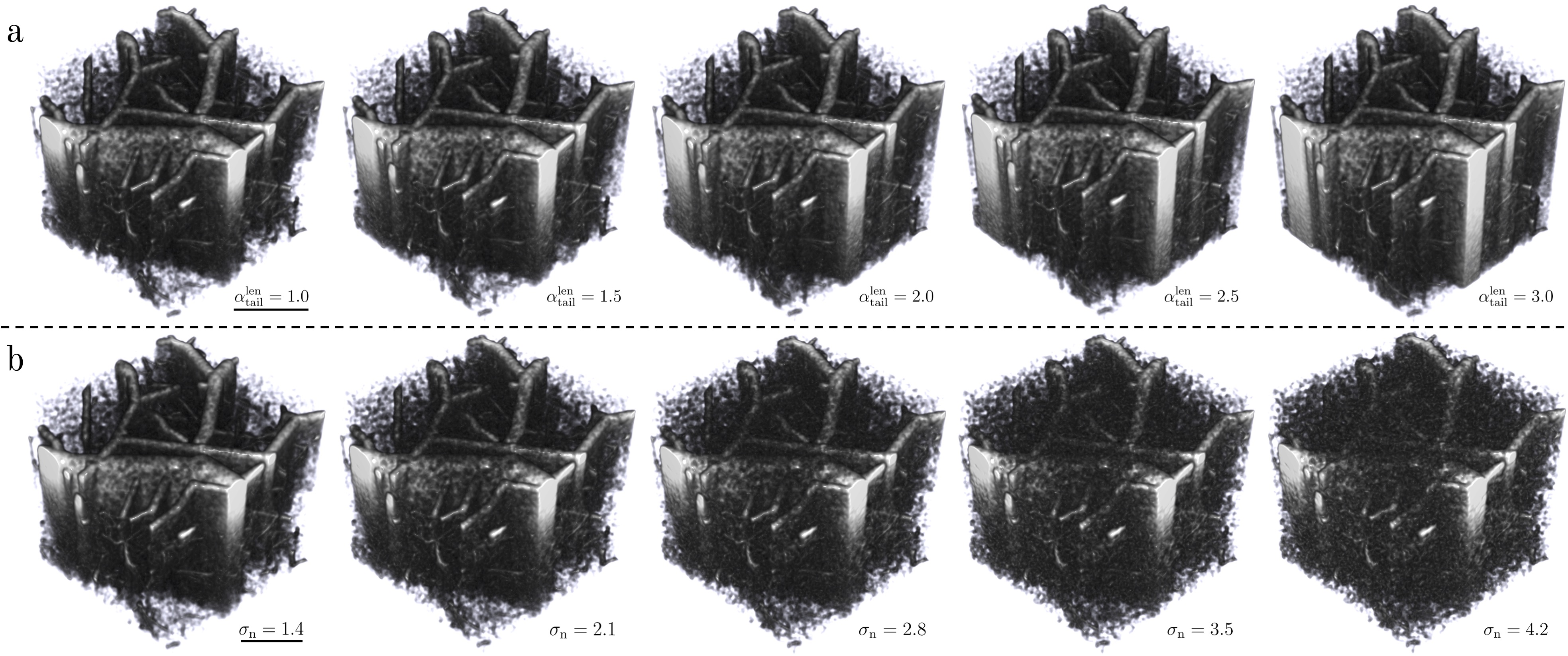}}
\caption{Effect of adjusting simulation parameters. The length of projection artifacts can, \emph{e.g.}, be increased to account for other beam geometries or the absence of external contrast agents by increasing the tail length factor $\alpha_{\text{tail}}^{\text{len}}$ (first row), while modifying $\sigma_{\text{n}}$ (second row) adjusts the intensity of local granular noise patterns. This flexibility enables us to cope with high variability in OCT system design and acquisition protocols.}
\label{fig:sim_params}
\end{figure}

\begin{figure}[h]
\centerline{\includegraphics[width=0.8\linewidth]{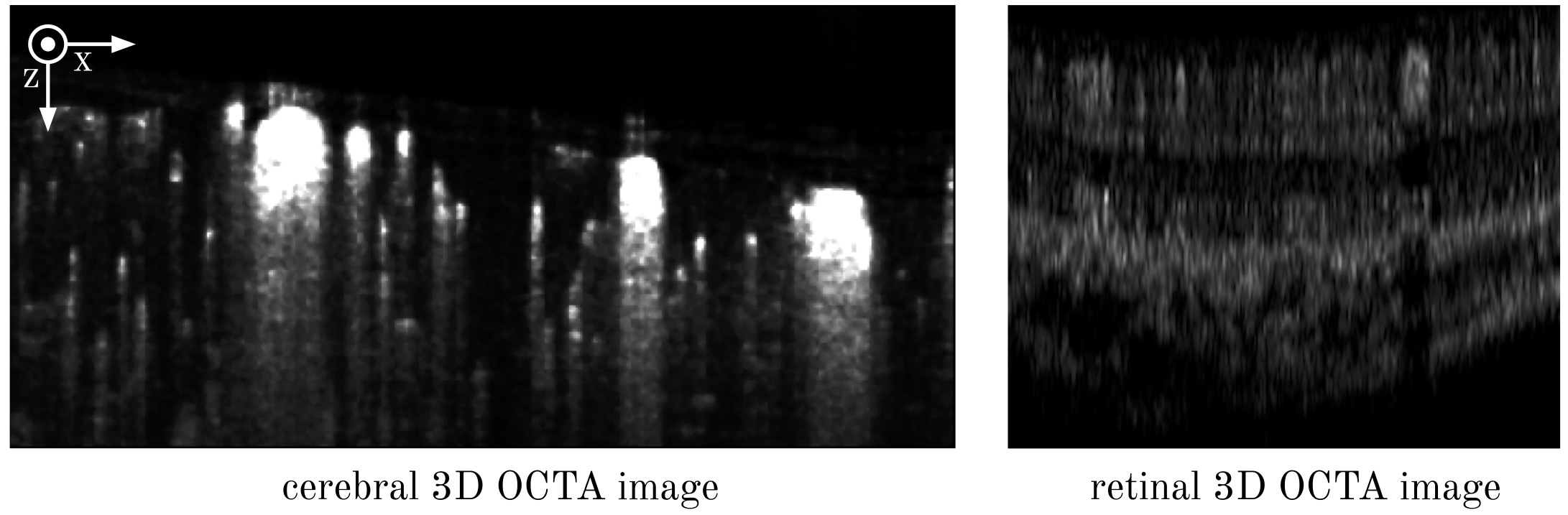}}
\caption{Slices of a cerebral and a retinal 3D OCTA image. The images differ not only drastically in underlying blood vessel morphology but also in general characteristics, such as signal-to-noise ratio, FOV, and voxel size. The depicted retinal 3D OCTA image originates from the OCTA-500 dataset~\cite{li2024octa}.}
\label{fig:retina_vs_cerebral}
\end{figure}

\begin{algorithm}[h]
\algrenewcommand{\algorithmiccomment}[1]{\hfill// #1}
\caption{Pseudocode: cerebral 3D OCTA artifact simulation}\label{alg:artifact_simulation}
\setstretch{1.0}
\begin{algorithmic}[1]
\scriptsize
\Require 
voxelized volume $I$,
metadata $I_{\text{meta}}$, 
upper radius threshold $r_{\text{max}}$,
radius threshold microvessels $r_{\text{micro}}$,
angle delta scaling factor $\gamma_{\Delta}$,
lambda intensity $\lambda_{\text{int}}$,
tail length factor $\alpha_{\text{tail}}^{\text{len}}$, 
tail intensity factor $\alpha_{\text{tail}}^{\text{int}}$,
tail noise mean $\mu_{\text{tail}}$,
tail noise std $\sigma_{\text{tail}}$,
granular noise mean $\mu_{\text{n}}$,
granular noise std $\sigma_{\text{n}}$,
smoothing sigma $\sigma_{\text{s}}$,
lower int. threshold $i_{\text{min}}$,
upper int. threshold $i_{\text{max}}$
\vspace{0.5em}

\State $I'  \gets \Call{ZerosLike}{I}$ \Comment{Initialize output volume}
\ForAll{\text{voxels} ($v_{\text{int}}$, $v_{\text{id}}$)  \textbf{in} $I$} \Comment{Loop over voxels given by intensity \& ID}
    \If{$v_{\text{int}}$ $\textbf{is}$ 0}
        \State $\textbf{continue}$ \Comment{Ignore empty space}
    \EndIf
    \State $\theta_{\text{z}}, r \gets I_{\text{meta}}[v_{\text{id}}]$ \Comment{Retrieve angle \& radius from metadata}
    \vspace{1em}

    \State{// Estimate intensity component contributed by vessel radius}
    \State $v_{\text{int}}^{\text{rad}} \gets \textproc{Scale}(\Call{Clip}{r, 0, r_{\text{max}}})$ \Comment{Clip with (0, $r_{\text{max}}$) \& scale to [0, 1]}
    \vspace{1em}

    \State{// Estimate intensity component contributed by angle between vessel \& z-axis}
    \State $v_{\text{int}}^{\text{micro}} \gets \textproc{Scale}(\Call{ExpDecay}{90^{\circ} - \theta_{\text{z}}})$. \Comment{Exponential signal decay \& scale to [0, 1]}
    \State $v_{\text{int}}^{\text{macro}} \gets \Call{Sigmoid}{\gamma_{\Delta} \cdot (r - r_{\text{micro}})}$ \Comment{Soft thresholding}
    \State $v_{\text{int}}^{\text{ang}} \gets \Call{Max}{v_{\text{int}}^{\text{micro}}, v_{\text{int}}^{\text{macro}}}$ \Comment{Signal decay just has an effect on microvessels}

    \vspace{1em}
    \State $v_{\text{int}} \gets \lambda_{\text{int}} \cdot v_{\text{int}}^{\text{ang}} + v_{\text{int}}^{\text{rad}}$ \Comment{Update voxel intensity}

    \vspace{1em}
    \If{$\Call{OccupancyBelow}{I, v_{\text{id}}}$ $\textbf{is not}$ 0} \Comment{Check if voxel not in lower vessel wall}
        \State $I'[v_{\text{id}}] \gets v_{\text{int}}$
        \State $\textbf{continue}$ \Comment{Do not model tail artifacts}
    \EndIf
    
    \vspace{1em}
    \State{// Model projection/tail artifacts}
    \State $l_{\text{tail}} \gets \alpha_{\text{tail}}^{\text{len}} \cdot v_{\text{int}}^{\text{rad}}$ \Comment{Determine tail length} 
    \State $t \gets \Call{GeomProg}{v_{\text{int}} \cdot \alpha_{\text{tail}}^{\text{int}}, 0, l_{\text{tail}}}$  \Comment{Model tail as sequence with geom. progression}
    \State $t \gets t + \Call{GaussianNoise}{\mu_{\text{tail}}, \sigma_{\text{tail}}}$  \Comment{Add random Gaussian noise to tails}
    \State $I' \gets I' + \Call{Clip}{t, 0, v_{\text{int}}}$  \Comment{Add clipped tail to output volume}
\EndFor

\vspace{1em}
\State{// Model local granular noise patterns}
\State $I' \gets I' + \Call{GaussianNoise}{\mu_{\text{n}}, \sigma_{\text{n}}}$
\State $I' \gets \Call{GaussianSmoothing}{I', \sigma_{\text{s}}}$
\State $I' \gets \textproc{Scale}(\Call{Clip}{I', i_{\text{min}}, i_{\text{max}}})$ \Comment{Clip with ($i_{\text{min}}, i_{\text{max}}$) \& scale to [0, 1]}

\State \Return $I'$ \Comment{Return synthetic cerebral 3D OCTA image}
\end{algorithmic}
\end{algorithm}

\end{document}